\documentclass{appolb}
\usepackage{epsfig}

\begin{document}
\title{ Implications of Mirror Dark Matter on Neutron Stars  %
  \thanks{Presented by Itzhak Goldman at the XXXV International Conference of Theoretical
Physics, "Matter to the Deepest", Ustron, Poland, September 12-18, 2011.
} 
}
\author{Itzhak Goldman
\address{ Department of Exact Sciences,  
Afeka Tel Aviv Academic Engineering College\\ Tel Aviv 69107, Israel } } 
 
\maketitle
\begin{abstract}
  We study the implications of asymmetric dark matter on neutron stars.  we construct a  "mixed neutron star" model composed of ordinary baryons and of  asymmetric dark matter baryons. We derive the general relativistic structure equations for each specie, the equation for the mass within  a given radius, and the redshift as function of radius. We present one specific numerical model
  as an illustrative example. In this example, the mass of the dark neutron equals half that of the ordinary neutron. The main results are: a total mass of  $3.74 M_{\odot}$, a
 total mass within the neutron-sphere equaling  $1.56 M_{\odot}$, the neutrons  mass is \  $1.34 M_{\odot}$,
  the   star radius is \  31.9    km, 
   the   neutron-sphere radius is \ 11.1 km, and the
      redshifts from the neutron-sphere and from the star surface are \  0.72, \    0.25, respectively.
We comment briefly on possible astrophysical implications. 
 
\end{abstract}
   
\section{Introduction}
\subsection{Background}

Cold dark matter   which is favored by most astrophysical and cosmological observations can be realized in symmetric and (or)  asymmetric scenarios.
In the first class of models, dark matter is made of stable $X$  particles and an equal amount of $\bar X$ antiparticles of mass $m_X$. In the early universe, these were   in thermal equilibrium and their residual abundance $\Omega_X$ is fixed at the "freeze-out" value when the rate of the Hubble expansion overcomes that of $\bar X-X $ annihilation rate. A prototypical example which have been most extensively studied is   SUSY(super-symmetry).  
Unfortunately, the searches for SUSY partners in the new large hadron collider (LHC)   have failed to detect them. 
More specifically, insofar as SUSY dark matter models are concerned searches for electrons, positrons or photons in clumped dark matter in and around our Galaxy, and for energetic neutrinos resulting from annihilations of Xs do not provide solid  "indirect" evidence for dark matter. Moreover, the ongoing direct underground searches put very strong bounds on the scattering crossections of massive Xs on nuclei. 

Additional constraints are related to accretion of galactic  SUSY WIMPS onto the sun. The increased density of the captured WIMPS, accelerates the rate of particle-antiparticle annihilation. 
 The resulting photons and electrons are   trapped in the sun but the resulting  ultra high energy  neutrinos are not.  Data from the ICE-CUBE Cerenkov radiation detector near the south pole,
severely constrain  such models  \cite{abbasi 09, abbasi 10} . 
 In the case of asymmetric dark matter, there will be no such annihilation in the sun. Moreover, once the fraction of dark matter particles in the sun exceeds the ratio of $\sigma_{Xn}/\sigma_{XX}$,   scattering on the already captured X particles in the sun dominates over scattering on the baryons.  
 
Finally,   the
observed number of satellite mini-halos around  the milky way galaxy  is two orders of magnitude smaller than predicted within the symmetric dark matter framework \cite{Klypin, Moore}.

Consequently there has been, in the very recent years, a renewed interest  in asymmetric dark matter  (ADM) which just like ordinary baryonic matter, is charge non-symmetric with say only the dark baryon (or generally only the particle)  excess remains after the annihilation of most antiparticles .While there is no single clear-cut explanation for the ordinary baryonic asymmetry, the required dark matter density is readily achieved if the mass ratio of the  X particles and baryons $m_X/m_b$ is tuned inversely with the corresponding ratio of asymmetries. An early example  of such a model has been proposed in \cite{nuss85}.  For recent   studies of asymmetric dark matter   see   \cite{An etal 2010,mrm11}     
 and references therein.

\subsection{Cosmological and astrophysical considerations}

The idea of asymmetric dark matter should reconcile with     astrophysical and cosmological data. An obvious constraint is imposed by big-bang nucleosynthesis, which implies that  the mirror neutrinos and  photons   do not contribute to the rate of the cosmic expansion at that era. Another
constraint is that the mirror large scale structure-formation should precede the recombination of   ordinary matter, in order to serve as seeds gravitational potential wells  for the latter. In the model described in  \cite{An etal 2010}  
the mirror neutrinos and the mirror photons are massive enough for these constraints to be satisfied. 
We note that even if
 the mirror neutrinos and photons are massless the above two constraints can be satisfied provided that  the  dark  cosmic-temperature is lower than
the ordinary one.

The details of large scale structure formation will depend on the specific model for the asymmetrical dark matter. Generally, being self interacting one may expect it to form mirror structures on all astrophysical   scales.  Also, asymmetric dark matter  will  no longer provide collisionless dark halos as in the symmetric cold dark matter case, so that  the flatness of galactic rotation curves will have to be readdressed in this new context.

\section{Composite "mixed neutron star"}
\subsection{Motivation and scope of present work}
  
  The mirror dark matter can form gravitationally bound structures. In particular they can form 
  "dark neutron stars". The latter can accrete ordinary matter and form composite neutron stars. 
One may contemplate additional formation scenarios for the composite neutron stars.
In this paper, we   focus on the structure of such a compact object and put aside   issues related to  and constraints following from formation scenarios, and the broader questions regarding the mirror astrophysics and cosmology.  This is an interim report on a work in progress \cite{dns}.

Such objects can have masses larger than those of ordinary neutron stars, while
otherwise having very similar observational signatures. The recent discovery of a  $2 M_{\odot}$  
binary radio pulsar \cite{2mo},  already severely constrains nuclear matter equations of state. A future observation of a neutron star  with a  mass exceeding $3 M_{\odot}$ would be very difficult to reconcile with an ordinary neutron star but will pose no problem for a mixed neutron star.

 The energy momentum tensor is taken to be that of two  ideal fluids
 \begin{equation}
\label{tmunu}
T^{\mu  \nu} = ( \rho_1 +p_1) u_1^{\mu}u_1^{\nu} - p_1 g^{\mu \nu} +( \rho_2 +p_2) u_2^{\mu}u_2^{\nu} - p_2 g^{\mu \nu}   
\end{equation} 
  Since any   inter-specie interaction must be weaker than the intra-specie interaction by a factor $\geq 10^{12}$,  each  energy momentum tensor is separately   conserved. In addition, the ordinary and dark baryon numbers are
separately conserved.
\subsection{Structure equations} 
We   look for a spherically symmetric static solution of a 2-fluid "mixed neutron star".   
 The line element squared of a spherically symmetric static metric can be written in the Schwarzshild coordinates $(t, r, \theta,\phi)$ as
 
\begin{equation}
\label{metric}
 ds^2=  e^{ 2\Phi(r)}c^2 dt^2 - e^{ 2\lambda(r)}dr^2 - r^2 \left(d \theta^2 + \sin^2(\theta) d\phi^2\right)
\end{equation}  
 The Einstein field equations  
that determine the metric, together with the separate covariant conservation laws can be shown  \cite{dns}) to lead to the following structure equations:

 \begin{equation}
\label{lambda } 
e^{- 2\lambda(r)}= \left( 1- 2\frac{G }{c^{2}}\frac{m(r)}{r}\right) 
\end{equation} 

where  $m(r)$ is the mass enclosed within   $r$   given by
 \begin{equation}
\label{mass} 
   m(r) = \int_0^r 4\pi \bigg(( \rho_1(r')  + \rho_2(r') \bigg) r'^2   dr'
\end{equation} 

For each of the species there exists  a hydrostatic equilibrium equation:

\begin{equation}
\label{equil1}
      \frac{d p_1(r)}{dr}= - G \bigg(\rho_1(r) + p_1(r)\bigg)\frac{m(r) + 4\pi r^3 \bigg(p_1(r)+ p_2(r)\bigg)}{r\bigg(r - 2 G c^{-2} m(r)\bigg)}
\end{equation} 

\begin{equation}
\label{equil2}
      \frac{d p_2(r)}{dr}= - G \bigg(\rho_2(r) + p_2(r)\bigg)\frac{m(r) + 4\pi r^3 \bigg(p_1(r)+ p_2(r)\bigg)}{r\bigg(r - 2 G c^{-2} m(r)\bigg)}
\end{equation} 
and $\phi(r)$ satisfies:
 \begin{equation}
\label{phi}
   \frac{d\Phi(r)}{dr}=  - \bigg(\rho_1(r) + p_1(r)\bigg)^{-1} \frac{d p_1(r)}{dr} =  - \bigg(\rho_2(r) + p_2(r)\bigg)^{-1} \frac{d p_2(r)}{dr}
\end{equation} 
 
The equations imply  that  each  fluid satisfies its own hydrostatic equilibrium equation  which is of the form of a  modified TOV  equation \cite{t, ov}. The two fluids are    coupled through $m(r)$ and  through the total pressure $p_1(r) + p_2(r)$.
 \subsection{Solution procedure}
 
 Given the two equations of state,  and 
  the two central energy densities,  the  
 TOV equations (\ref{equil1}, \ref{equil2}) are integrated up to $r=R_1$ where $p_1(R_1) = 0$. Specie 1 is confined within this radius. From this radius on,  only TOV2 (equation \ref{equil2}) is integrated until the radius $R_2$ where  $p_2(R_2) = 0$. This is the star radius.
 Once the solution is obtained, equation (\ref{phi}) is solved with the boundary condition
 $\Phi(R_2)=\frac{1}{2} \ln { \left( 1- 2\frac{G }{c^{2}}\frac{m}{R_2}\right)}$, with
 $m= m(R_2)$ being the mass of the mixed neutron star.
 
In this way, a two-parameters (the central densities) family of static models is obtained, in contrast with the ordinary neutron star models that form a one-parameter (the central density) family.  
 
    \section{An illustrative example }
 
\begin{figure}[h]
\begin{center} $
\begin{array}{cc}
\includegraphics[scale=0.55]{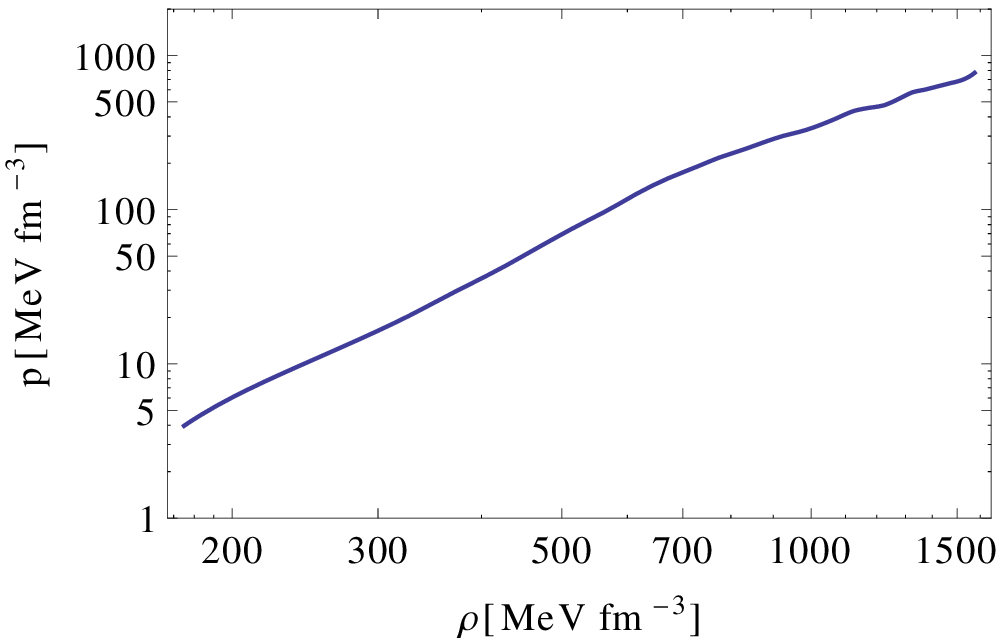}  &
\includegraphics[scale=0.4]{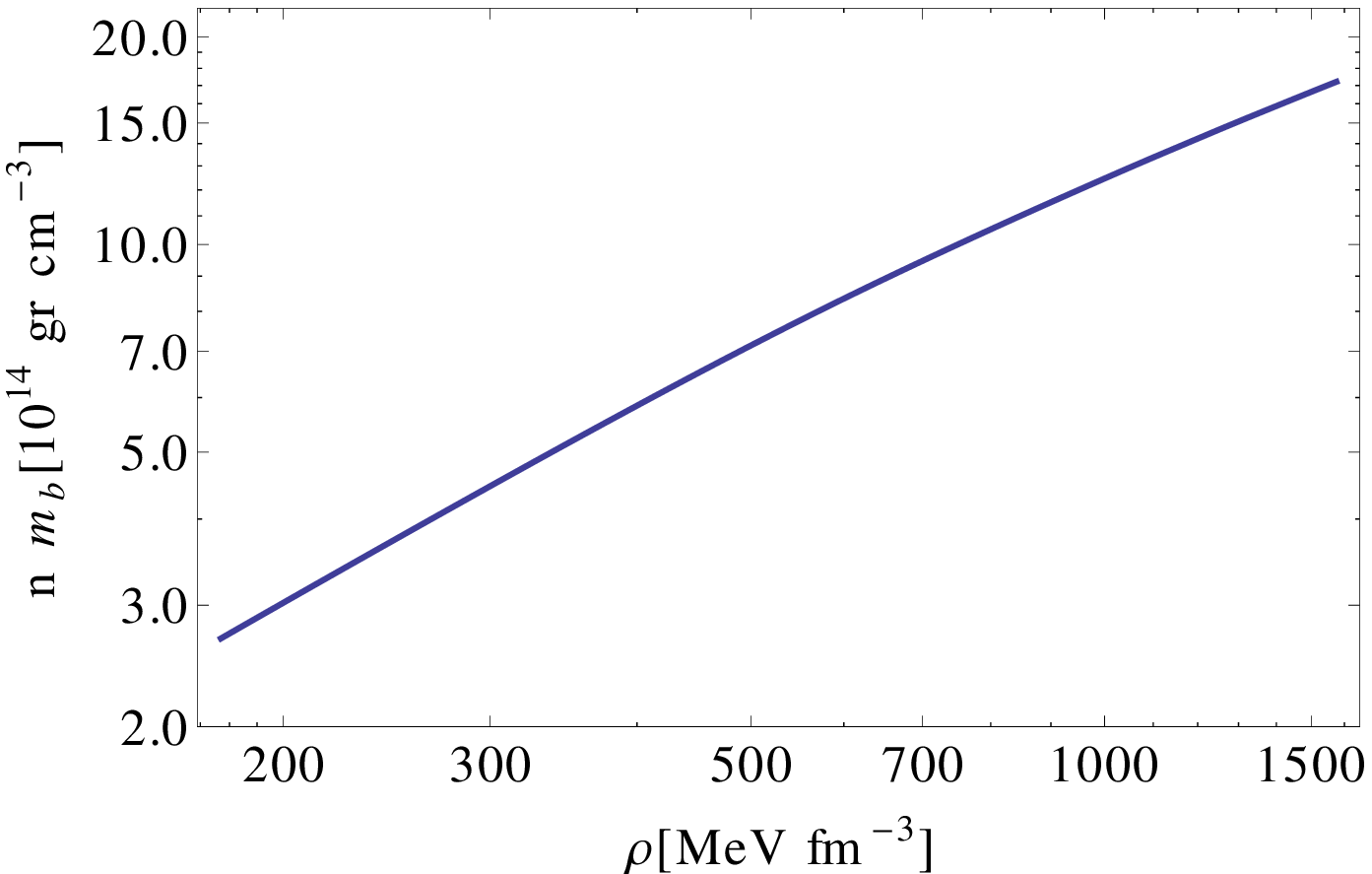} 
\end{array}$
\end{center} 
\caption{{\bf left }: pressure as function of energy density.  {\bf right}: number density as function of energy density.} 
 \label{Fig:pn_rho}
\end{figure}

  To illustrate the  model characteristics we present here a generic example.  
We employ a nuclear matter equation of state 
  that was obtained by fitting observational data of x-ray bursters \cite{steiner}.  Fig.~\ref{Fig:pn_rho} displays
  the pressure and the baryon number as functions of the energy density.  
  The maximal  ordinary neutron star mass for this equation of state is $2.44 M_{\odot}$ and the  corresponding radius is $11.7 \ km$.

For the dark baryons we use the {\it same} equation of state as for the neutrons with the appropriate scaling, taking into account   baryons masses ratio. This choice seems to be the minimal and simplest one. 

The ordinary neutrons star masses satisfy $m\simeq m_{pl}^3m_b^{-2}$ \cite{dns} with $ m_{pl}$ denoting the Planck mass. Therefore,
in order to obtain a mixed neutron star with mass larger than ordinary neutron stars, it is required that $m_D<m_b$ where $m_D$ is the dark baryon mass.   In this example we chose $m_D= 0.5 m_b$, so that  the dark equation of state is

 \begin{equation}
 \label{p2rho2}
   p_2(\rho_2)=\frac{1}{16}  p_1 (16 \rho_2 )  
 \end{equation}
  
The  mass of a  pure dark neutron star   will be   $\sim 8M_{\odot}$ and the corresponding radius   is $ \sim 50 km$.  It is expected that  the mixed neutron star solution  would yield    a   mass, and radius  intermediate   between those of a neutron star and a pure dark neutron star.
 
 The values of the central energy densities   in this example are:
 \begin{equation}
 \label{rho120}
 \rho_1 (0)= 600\ {\rm Mev\ fm}^{-3}  \ ,\ \ \  \rho_2 (0)= \frac{1300}{16}    \  {\rm Mev\ fm}^{-3} 
 \end{equation} 
   
 \section{Computation  results}
\begin{figure}[h]
\begin{center} $
\begin{array}{cc}
\includegraphics[scale=0.4]{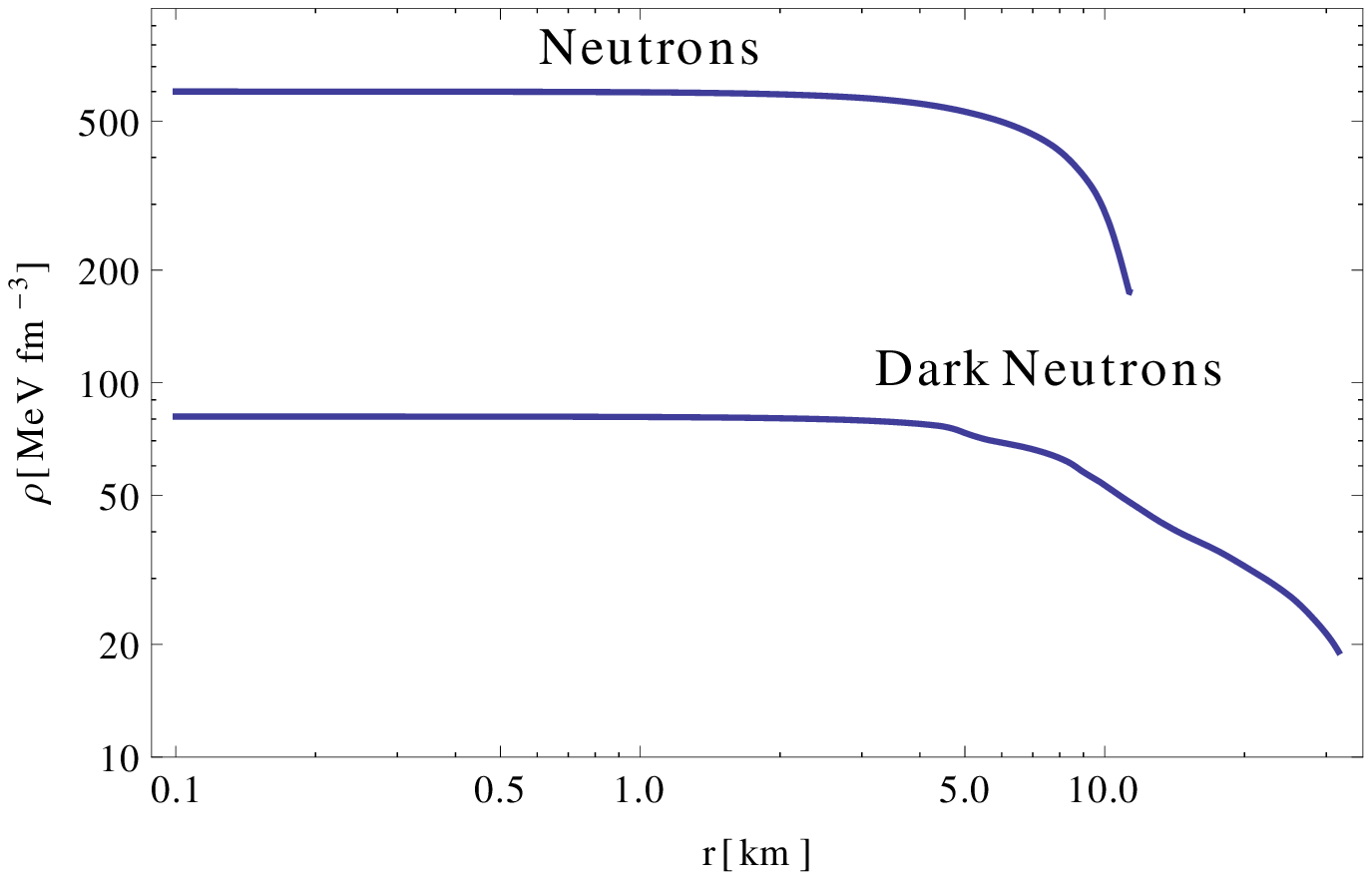}   &
 \includegraphics[scale=0.4]{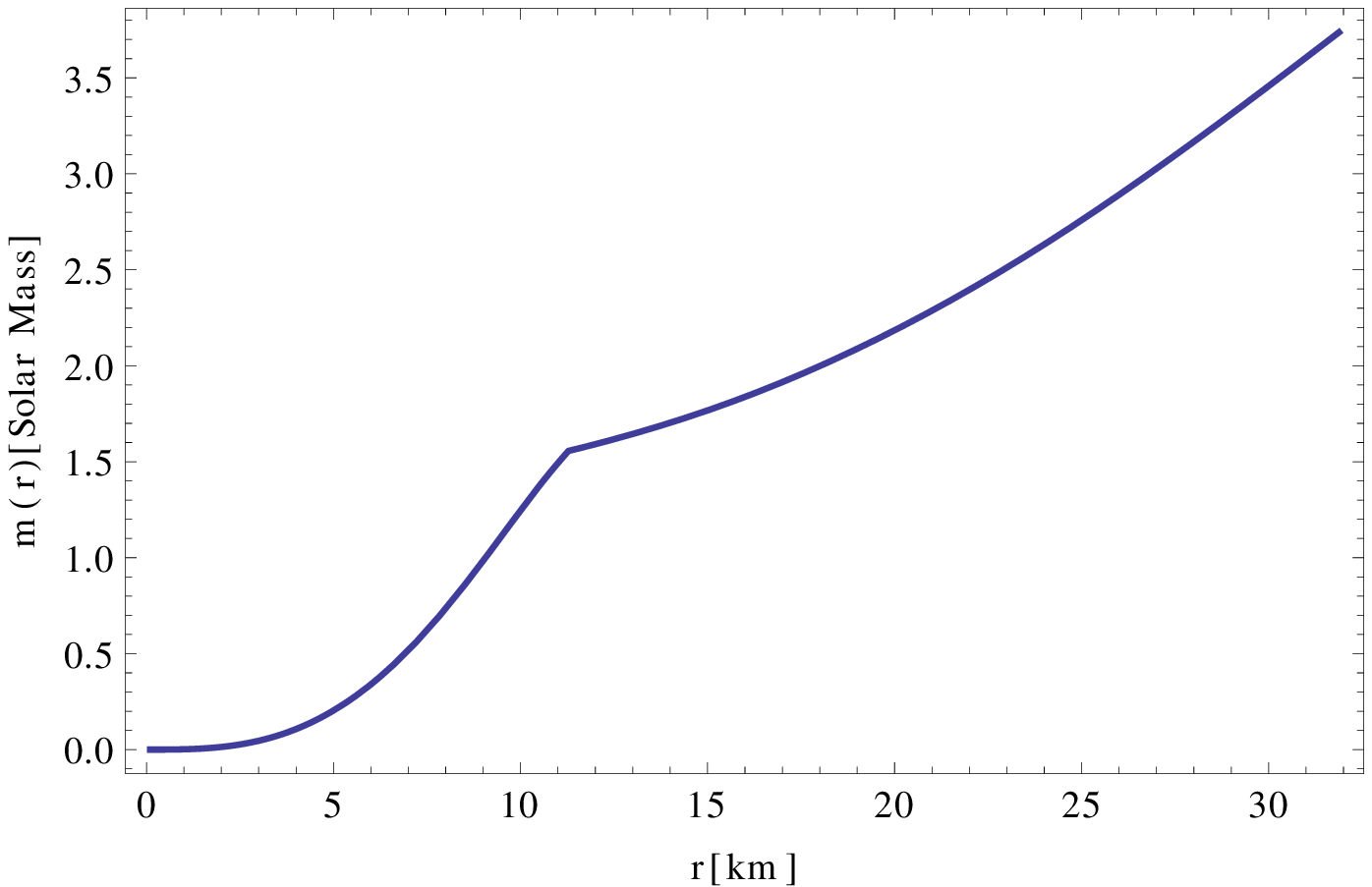}
\end{array}$
\end{center} 
\caption{ energy densities (left) and enclosed mass (right)  as function of radius. } 
 \label{Fig:rho12m}
 \end{figure}

\begin{figure}[h] 
\centerline{\includegraphics[scale=0.44]{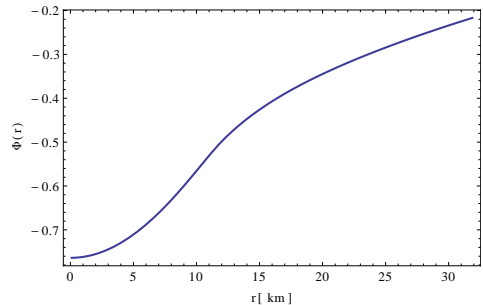}}
 \caption{$  \Phi $ as function of radius.}
   \label{phi_r}
\end{figure}

 The computation results are summarized in Table \ref{tbl}. The first row contains the star total mass, the dark mass, the neutrons mass, and the total mass within the neutron-sphere. The second row specifies the star radius, the neutron-sphere radius, the redshifts for these two radii, and the neutrons binding energy.   The  detailed r-dependence of the energy densities, the enclosed mass $m(r)$, and $\Phi(r)$ are displayed in figures (\ref{Fig:rho12m}) and  (\ref{phi_r}), respectively.
\begin{table}[h]
\caption{Model Results}
\label{tbl}
     \vskip 0.5 truecm 
\begin{center} 
\begin{tabular} {  p{1.6 cm}     p{1.6  cm} p{4.6cm}     p{2.2 cm}   } 
 \hline  \hline
m & $m_{dark}$  & $m_{neutrons}$ & $m(R_1)$
 \\   \hline
   $3.74 M_{\odot}$& $2.4 M_{\odot}$  & $1.34 M_{\odot}$ &   $1.56 M_{\odot}$\\
    \\ \hline  \hline
 $R_2$&
  $R_1$  & Redshifts & Neutrons BE  \\
  \hline   
      31.9    km &  11.1 km    &   $z(R_1)$=0.72,\   $z(R_2)$=0.25    &  22\%
 
   \end{tabular} 
   \end{center}
 \end{table} 
  
\section{Discussion}
We have demonstrated    that a   mixed neutron star can, as expected, have
a mass higher than ordinary neutron stars. At the same time the physical  radius, as probed by ordinary massless and massive particles, is the neutron-sphere radius which is similar   to
the radius of ordinary neutron stars.

An important question, not addressed  here, is that of stability.   Since the models form a two-parameters family (the central densities) the question of stability is more complex than in the one parameter family of ordinary neutron star models.  
There are a number of quite interesting astrophysical implications with regard to phenomenology of compact x-ray sources. Can some of the stellar mass binary  black holes be actually mixed neutron stars? The  neutron-sphere redshift is about  50\% higher than in the ordinary neutron star case, which may have interesting results for the temperature, radius and luminosity measured by a distant observer.  The larger neutrons binding energy would lead to a smaller value of the maximal neutrons mass, compared to an ordinary neutron star.

 I wish to thank my collaborators, R. Mohapatra, S. Nussinov, D. Teplitz and V.~Teplitz  
as well as  W. Kluzniak and L. Zdunik for interesting discussions. Thanks are due  to   the Afeka
College Research Committee  for financial support.

\end{document}